\documentclass[12pt]{article}
\usepackage{amssymb}

\input{tcilatex}

\begin{document}

\bigskip \baselineskip0.8cm \textwidth16.5 cm

\begin{center}
\textbf{The Critical Finite Size Scaling Relation of the Order-Parameter
Probability Distribution for the Three-Dimensional Ising Model on the Creutz
Cellular Automaton }

B. Kutlu$^{1}$ and M.Civi$^{2}$

$^{1,2}$Gazi \"{U}niversitesi, Fen-Edebiyat Fak\"{u}ltesi , Fizik Anabilim
Dal\i , Ankara, Turkey

$^{1}$e-mail: bkutlu@gazi.edu.tr

$^{2}$e-mail: mcivi@gazi.edu.tr
\end{center}

We study the order parameter probability distribution at the critical point
for the three-dimensional spin-1/2 and spin-1 Ising models on the simple
cubic lattice with periodic boundary conditions. The finite size scaling
relation for the order parameter probability distribution is tested and
verified numerically by microcanonical Creutz cellular automata simulations.
The state critical exponent $\delta $, which characteries the far tail
regime of the scaling order parameter probability distribution, is estimated
for 3-d Ising models using the cellular automaton simulations at the
critical temperature. The results are in good agreement with the monte carlo
calculations.


A quantity of central importance for the finite-size scaling analyses of the
critical phenomena is the order parameter probability distribution $P(M)$.
The knowledge of \ the finite size scaling function for $P(M)$ of the Ising
model makes it possible to calculate all the moments of the order parameter
and all its cumulants.$^{[1-5]}$ Most properties of $P(M)$ are known from
computer simulations.$^{[6-14]}$ For d=3, the corresponding finite size
scaling function do not exist in the analytical form. There are effort to
get simple analytical functions by fitting to Monte Carlo results at the
critical temperature for the infinite lattice.$^{[10-13]}$ The main purpose
of these study is to test the finite size scaling relations for the order
parameter probability distribution of the three-dimensional spin-1/2 and
spin-1 Ising models by the microcanonical Creutz cellular automaton
algorithm on the simple cubic lattice and to obtain the value of the state
critical exponent $\delta $ . The Creutz cellular automaton (CCA) for the
Ising model has been proven to be successful in producing the values of the
universal statical critical exponents and the critical temperature in two
and higher dimensions.$^{[15-20]}$ The CCA\ algorithm, which was first
introduced by Creutz,$^{[20]}$ is a microcanonical algorithm interpolating
between the canonical Monte Carlo and Molecular dynamics techniques.

In this paper, the probability distribution of order parameter is obtained
for the two variants of the Ising model on the CCA. The first model is the
3-d spin-1/2 Ising model on the simple cubic lattice. For the zero external
magnetic field ($H=0$), the Hamiltonian of the model is given by

\begin{equation}
H_{I}=-J\sum_{<ij>}S_{i}S_{j}
\end{equation}%
where $S_{i}=\pm 1$ and the sum is carried out over all nearest-neighboring
(nn) spin pairs. The parameter $J$ ( $J>0$ ) is the ferromagnetic coupling
constant. The simulations are carried out on simple cubic lattice $L$x$L$x$L$
of linear dimensions $L=$ 16, 18, 20, 24 and 40 with periodic boundary
conditions. The second model is the Blume-Capel(BC) model without single-ion
anisotropy parameter ( $D=0$ ). The BC model is a spin-1 Ising model.$%
^{[21,22]}$ It has the same Hamiltonian with spin-1/2 Ising model for the $%
H=0$ . Here, the spins can take three discrete values -1, 0 and 1. The BC
model is simulated using an improved algorithm$^{[18]}$ from CCA\ for simple
cubic lattice $L$x$L$x$L$ of linear dimensions $L=$ 16, 20, 24, 28 and 32
with periodic boundary conditions. The data are averages over the lattice
and the number of time steps (1.000.000) during which the cellular automaton
develops. The simulations are done $20$ times with different initial
configurations at the critical point for the Ising models.

The infinite lattice critical temperature values for spin-1/2 and spin-1
Ising models are estimated from the temperature variation of the Binder
forth-order cumulant for the finite lattices. The Binder forth-order
cumulant of the order parameter, which is used to estimate the infinite
lattice critical temperature, is given by

\begin{equation}
g_{L}=1-\frac{\langle M^{4}\rangle }{3\langle M^{2}\rangle ^{2}}
\end{equation}%
The temperature variations of \ the Binder cumulant are illustrated in Fig.1
(a) for spin-1/2 and (b) in Fig.1(b) for spin-1 Ising model. The infinite
lattice critical temperatures are obtained from the intersection of the
finite lattice Binder cumulant curves. The critical temperature value for
spin-1/2 Ising model is estimated as $kT_{c}/J=4.511\pm 0.002$ which is in
good agreement with Monte Carlo calculations.$^{[23-26]}$ For the spin-1
Ising model , the estimated critical temperature ($kT_{c}/J=3.197\pm 0.002$)
is also in a good agreement with a series expansion results.$^{[27-29]}$ The
behavior of the order parameter probability distribution at the critical
point is investigated at these estimated critical temperatures.

The order parameter probability distribution $P_{L}(M)$ is calculated by

\begin{equation}
P_{L}(M)=\frac{N_{M}}{N_{CCAS}}
\end{equation}%
where $N_{M}$ is the number of times that magnetization $M$ appears, and $%
N_{CCAS}$ is total number of Creutz cellular automaton steps. Histograms of
200 bins are used in plotting $P_{L}(M)$ at critical point for finite
lattice sizes which are shown in Fig.2 (a) for spin-1/2 and in Fig.2(b) for
spin-1 Ising model. The simulations have been performed for the twenty
different initial configurations with a constant total energy at critical
point. The data of $P_{L}(M)$ are obtained from averages of data for
different initial configurations. The relative standard deviation (RSD) of
the average $P_{L}(M)$ values is approximately in the interval $5\%-10\%$.

The finite size scaling hypothesis for the probability distribution of order
parameter for Ising model can be expressed generally as follows$^{[8]}$

\begin{equation}
P_{L}(M)=bP^{\ast }(M^{\ast })
\end{equation}%
where $b=b_{0}L^{\beta /\nu }$ , $\beta $ and $\nu $ are critical exponents, 
$M^{\ast }=bM$, $b_{0}$ is a constant, and $P^{\ast }(M^{\ast })$ is a
universal scaling function. To compute the normalized distribution $P^{\ast
}(M^{\ast })$ via Eq.4 one has evaluate the pre-factor $b$. The value of $b$
for each lattice sizes can be easily calculated by

\begin{equation}
b=1/(\left\langle M^{2}\right\rangle -\left\langle M\right\rangle ^{2})^{1/2}
\end{equation}%
Thus, we used the Eq.4 for obtained of the universal function $P^{\ast
}(M^{\ast })$ as in Ref.[8]. At the critical point, the plots of finite size
scaling for $P(M)$ are illustrated in Fig.3 (a) for spin-1/2 and in Fig.
3(b) for spin-1 models. The microcanonical simulations have been done on
simple cubic lattices at $kT_{c}/J=4.511$ for spin-1/2 and at $%
kT_{c}/J=3.197 $ for spin-1. For both models, the universality at critical
point can be easily seen from Fig.3(a) and (b). The scaling probability
distributions $P^{\ast }(M^{\ast })$ for spin-1/2 and spin-1 Ising models
coincided with each other for all $\ M^{\ast }$ values. These results verify
the finite size scaling relation given in Eq. 4 for the order parameter
probability distribution of the three dimensional Ising model. On the other
hand, the log-log plots of the pre-factor $b$ against $L$ are shown in
Fig.4. The slopes of the data line a single curve are given the values of $%
\beta /\nu $=0.498$\pm 0.003$ for spin-1/2 and $\beta /\nu $=0.492$\pm 0.003$
for spin-1. Furthermore, the $\beta /\nu $ values for both models are
estimated using the scaling relation of the order parameter at the critical
temperature[Fig.4(b)]. The slopes of the log-log plots of $M(T_{c})$
against\ to $L$ are given the values the $\beta /\nu $=0.50$\pm 0.01$ for
spin-1/2 and $\beta /\nu $=0.51$\pm 0.01$ for spin-1. All estimated values
of $\beta /\nu $ are in good agreement with universal value$^{[5,30]}$.

On the other hand , for the $\left\vert M^{\ast }\right\vert >1$ and the
periodic boundary conditions, the scaling function of the order parameter
probability distribution is expected to have the following exponential form$%
^{[13]}$ 
\begin{equation}
P^{\ast }(M^{\ast })\propto \exp \left( -AM^{\ast \text{ }\delta +1}\right) 
\end{equation}%
where $\delta $ denotes the equation of state exponent. Its value can be
obtained from the log-log plot of $lnP^{\ast }(M^{\ast })$ against to $%
M^{\ast }$. In the far tail regions, the slopes of data line a single curve
gives approximately $\delta +1=5.8$ for spin-1/2 and $\delta +1=5.7$ for
spin-1 on the right and left tails. The behavior of tails for large $M^{\ast
}$ are shown in Fig.5 (a) and (b) for spin models. The estimated values of $%
\delta +1$ are in good agreement with Monte Carlo results $(\delta =4.8)$ $%
^{[6,13]}$ for 3-d Ising model. \ \ Finally, \ the finite size scaling
relation for the order parameter probability distributions of the Ising
models is verified by Creutz cellular automaton simulations numerically.

\textbf{Acknowledgements}

This work is supported by a grant from Gazi University (BAP:05/2003-07).

\textbf{References}

[1] Binder K\ 1981 Phys.Rev. Lett. 47 693

[2] Bruce A D 1985 .J.Phys. A.:Math.Gen. 18 L873

[3] Eisenriegler E and Tomaschitz R \ 1987 Phys. Rev. B. 35 4876

[4] Binder K, Nauenberg M, Privman V and Young A P 1985 Phys. Rev. B 31 1498

[5] Privman V 1990 Finite Size Scaling and Numerical Simulation of
Statistical Systems (World Scientific, Singapore)

[6] Bruce A D 1995 .J. Phys. A.: Math.Gen. 28 3345

[7] Martinos S S, Malakis A and Hadjiagapiou I 2004 Physica A. 331 182

[8] Martins P H L and Plascak J A 2004 Braz. J. Phys. 34 433

[9] Berg B A, Billoire A and Janke W 2002 Phys. Rev. E 66 046122

[10] Bl\"{o}te H W J, Heringa J R and Tsypin M M 2000 Phys. Rev. E 62 77

[11] Tsypin M M and Bl\"{o}te H W 2000 Phys. Rev. E 62 73

[12] Rudnick J, Lay W and Jasnow D1998 Phys.Rev. E 58, 2902

[13] Hilfer R and Wilding N B 1995 J. Phys.\ A: Math. Gen. 28 L281

[14] Aktekin N 2004 G.\"{U}. J. Science 17 59

[15] \"{O}zkan A, Sefero\u{g}lu N and Kutlu B 2006 Physica A 362 327

[16] Kutlu B, \"{O}zkan A, Sefero\u{g}lu N, Solak A and Binal B 2005 Int. J.
Mod. Phys.C 16 933

[17] Kutlu B 2003 Int. J. Mod. Phys. C 14 1305; 2001 Int. J. Mod. Phys. C 12
1401

[18] Aktekin N 2000 Ann.. Rev. Comp. Phys. VII ed Stauffer D ( World
Scientific, Singapore) p1

[19] Merdan Z, G\"{u}nen A and M\"{u}laz\i mo\u{g}lu G 2005 Int. J. Mod.
Phys. C 16 1269; Merdan Z and Bay\i rl\i\ M\ 2005 App. Math. and Comp. 167
212; Merdan Z and Erdem R 2004 Phy. Lett. A 330 403; Merdan Z,G\"{u}nen A
and \c{C}avdar \c{S} 2006 Physica A 359 415

[20] Creutz M 1986 Ann.Phys.167 62

[21] Blume M B 1966 Phys. Rev. B 141 517

[22] Capel H W 1966 Physica(Utrecht) 32 966

[23] Salmon Z and Adler J 1998 Int. J. Mod. Phys. C 9 195

[24] Bl\"{o}te H W J, Luijten E and Heringa J R 1995 J. Phys. A28 6289

[25] Talapov A L and Bl\"{o}te H W J 1996 J.Phys.A 29 5727

[26] Baillic C F, Gupta R, Hawick K A and Pawley G S 1992 Phys. Rev. B45
10438

[27] Saul D M, Wortis M and Stauffer D 1974 Phys.Rev. B 9 4964

[28] Brankov J G, Przystawa J and Pravecki E 1972 J.Phys.C 5 3384

[29] Du A, Y\"{u} Y Q and Liu \ H J 2003 Physica A 320 387

[30] Janke W and Villanova R 2002 Phys. Rev. B, 66, 134208

\textbf{Figure Captions}

Fig.1 The variations of Binder cumulant against to $kT/J$ for (a) spin-1/2
and (b).spin-1 Ising model.

Fig.2.The order parameter probability distribution at the critical point for
(a) spin-1/2 and (b).spin-1 Ising model on the simple cubic lattices.
Simulations were performed at $kT_{C}/J=4.511$ for spin-1/2 and at $%
kT_{C}/J=3.197$ for spin-1 model.

Fig.3. Scaling functions $P^{\ast }(M^{\ast })$ for\ (a) three-dimensional
spin-1/2 and (b) spin-1 Ising model on simple cubic lattices.

Fig.4.(a ) The Log-log plot of b against to L for spin-1/2 and spin-1 Ising
models. \ (b) The log-log plots of $M(T_{c})$ against\ to $L$ . The values
of slopes are in good agreement with the universal $\beta /\nu $ value.

Fig.5. The log-log plots of ln( $P^{\ast }(M^{\ast })$ ) against to $M^{\ast
}$ for (a) spin-1/2 and (b) spin-1 Ising model. The slopes are equal to $%
\delta +1=5.8$ for spin-1/2 and $\delta +1=5.7$ for spin-1 model in the far
tail regions$.$

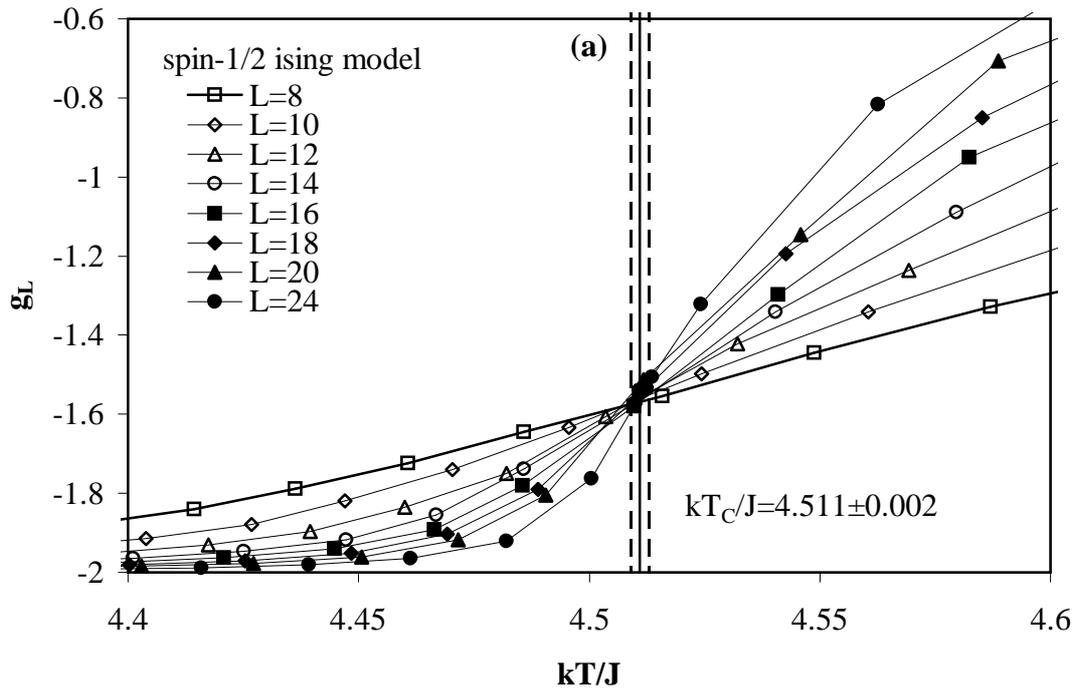
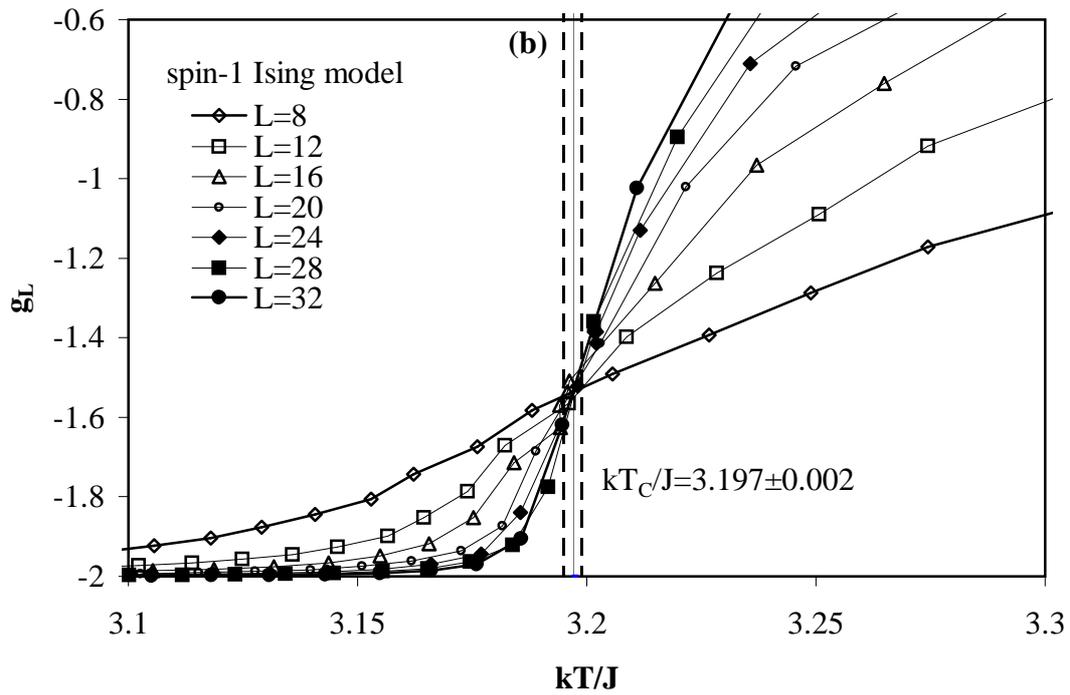
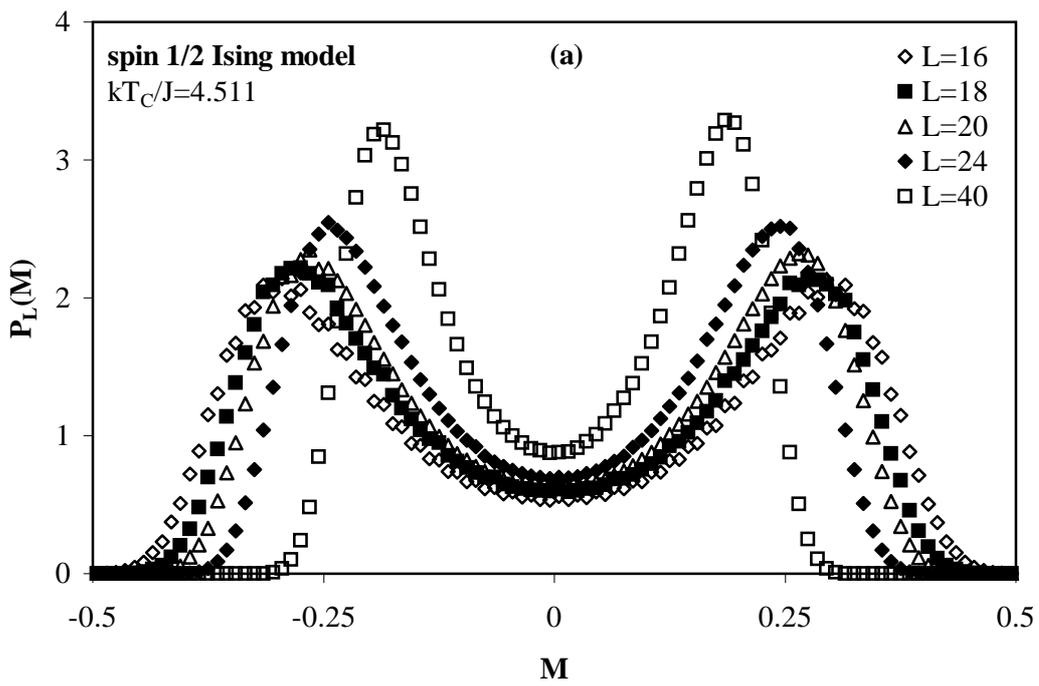
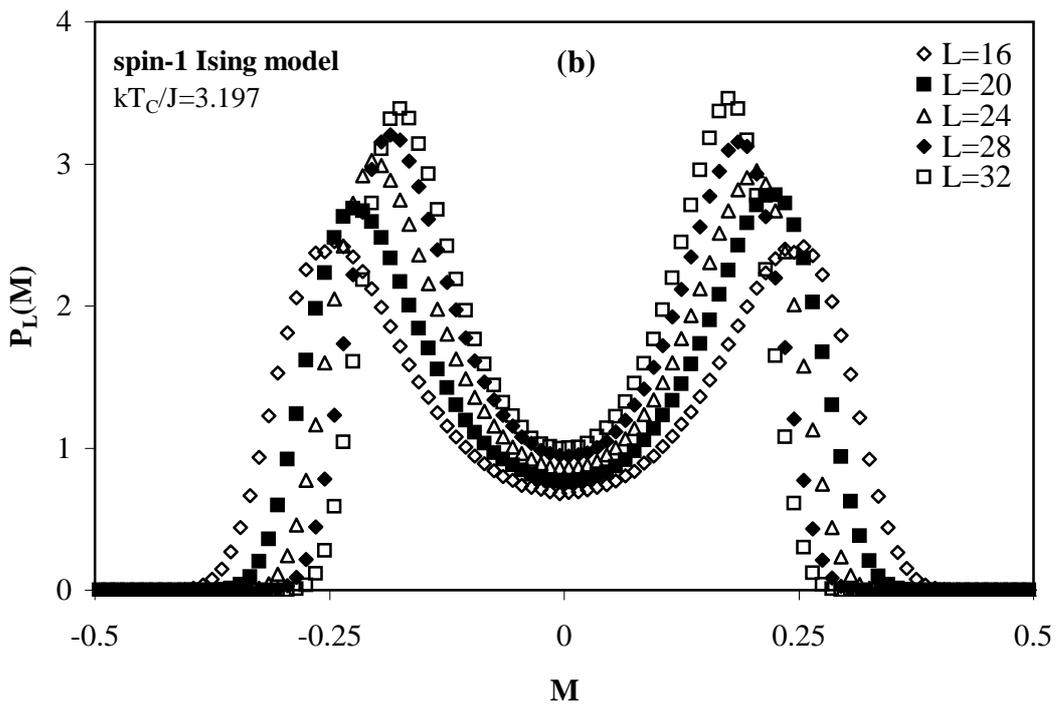
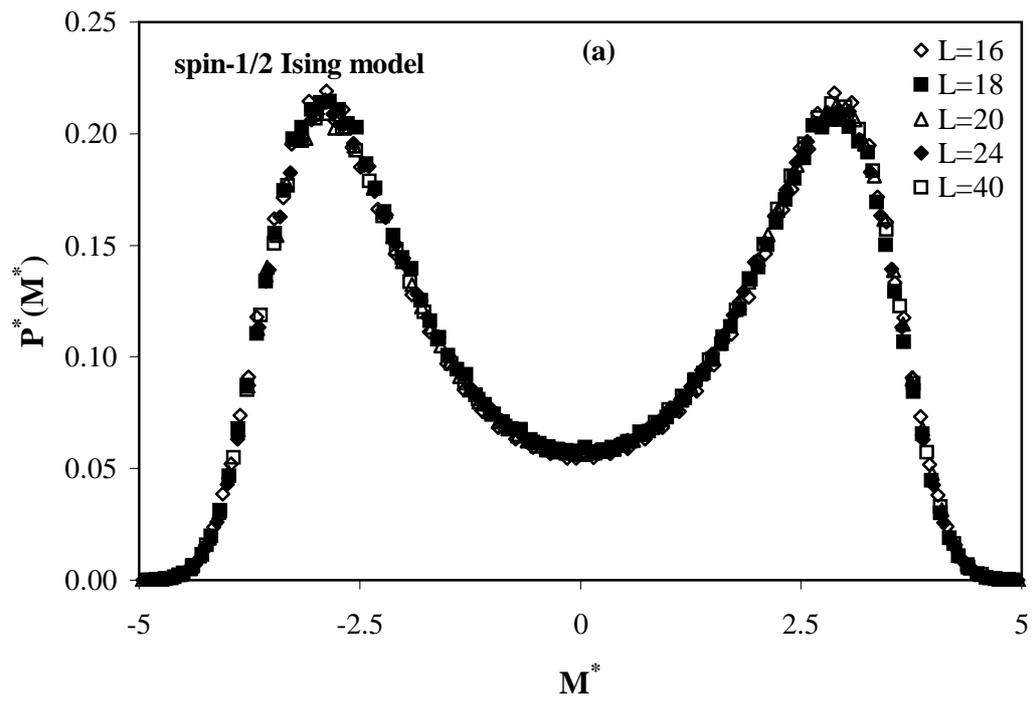
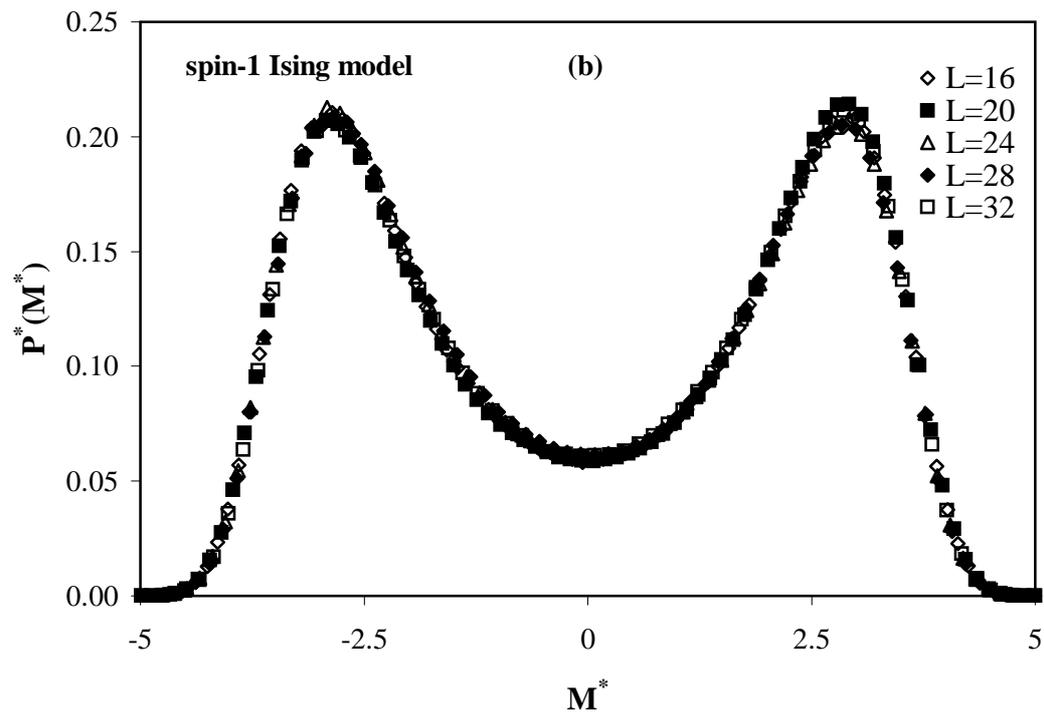
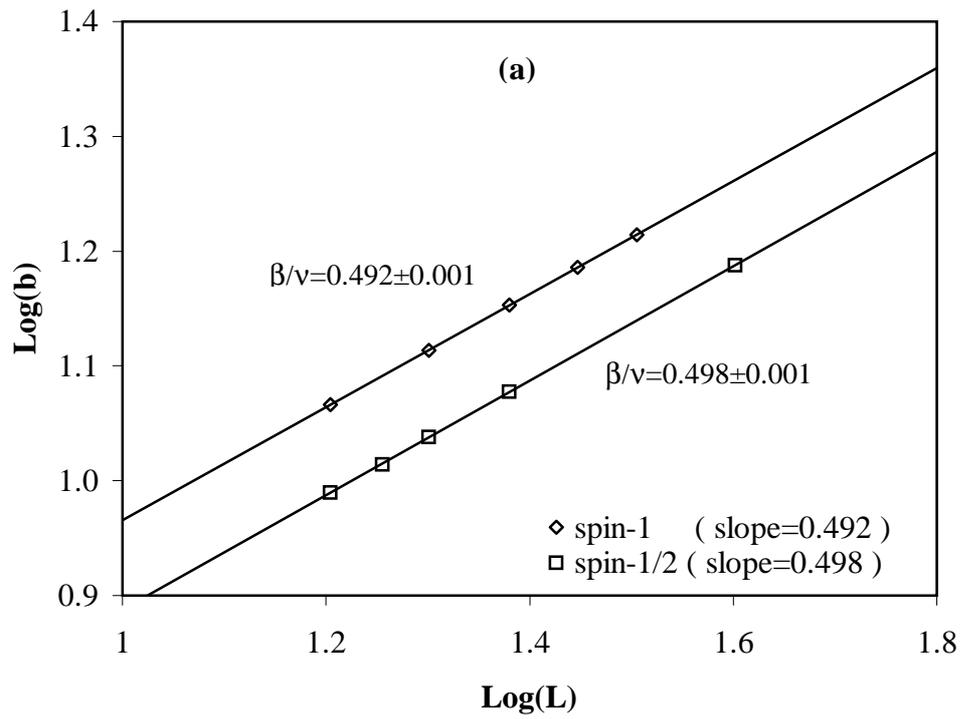
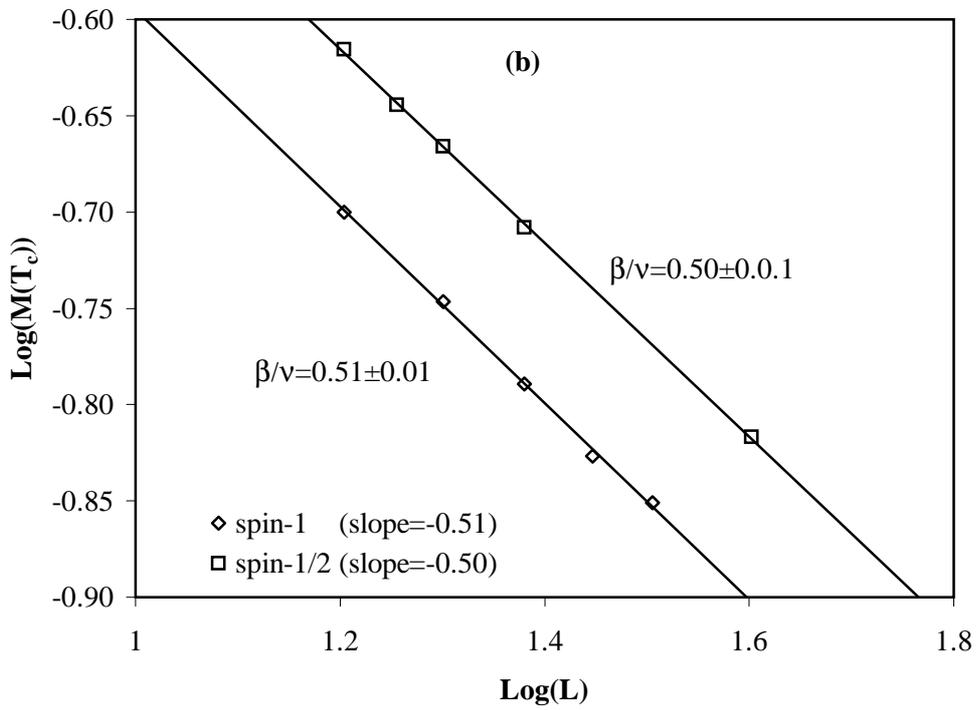
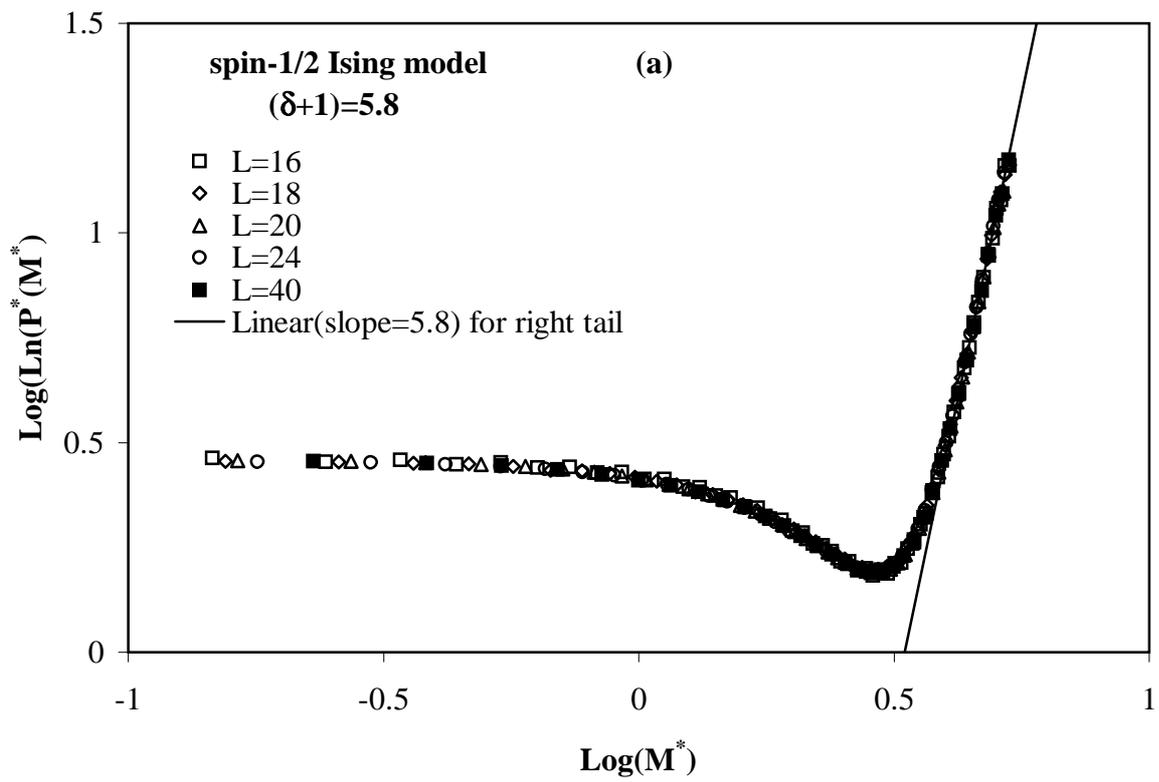
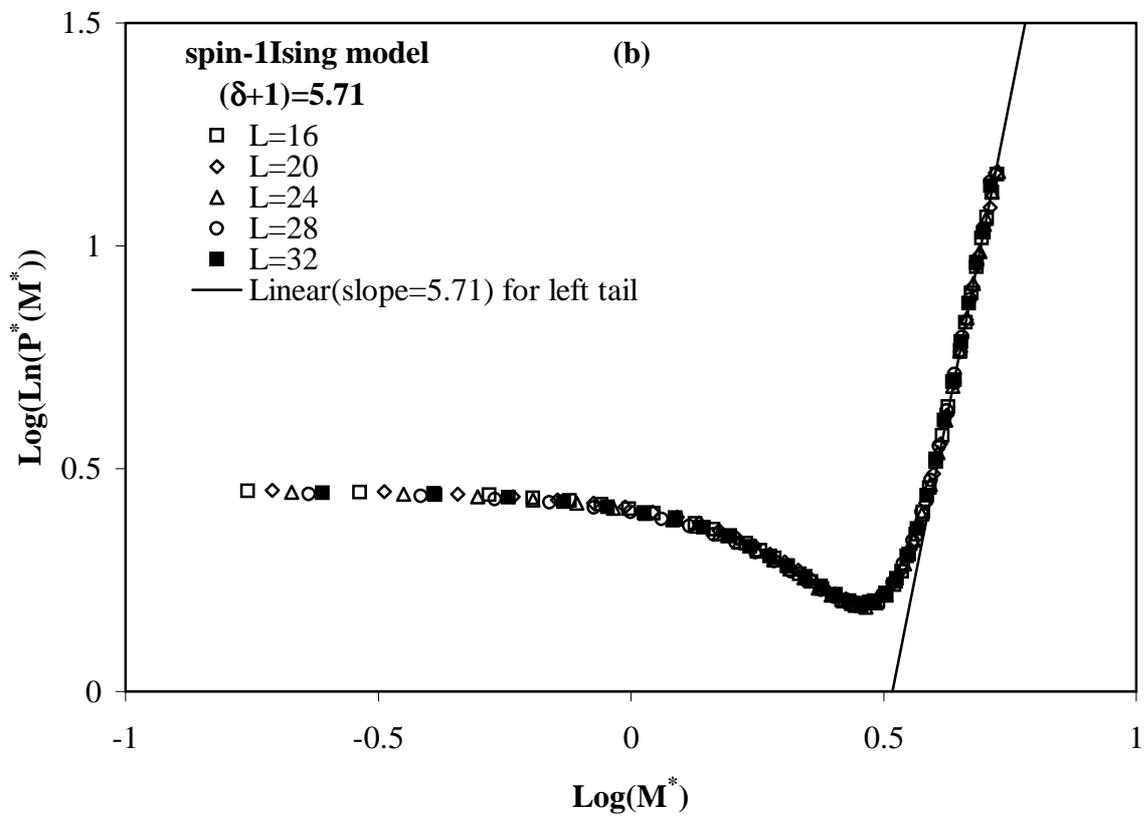
\end{document}